# THz excitations in α-RuCl₃:
# Majorana fermions, rigid-plane shear and compression modes

S. Reschke,[1] V. Tsurkan,[1,2] S.-H. Do,[3] K.-Y. Choi,[3] P. Lunkenheimer,[1] Zhe Wang,[4] and A. Loidl[1*]

[1]*Experimental Physics V, Center for Electronic Correlations and Magnetism, University of Augsburg, 86159 Augsburg, Germany*
[2]*Institute of Applied Physics, MD 2028 Chisinau, Republic of Moldova*
[3]*Department of Physics, Chung-Ang University, Seoul 06974, Republic of Korea*
[4]*Institute of Radiation Physics, Helmholtz-Zentrum Dresden-Rossendorf, 01328 Dresden, Germany*

Spin liquids may host emergent quasiparticles, collective excitations of the spin degrees of freedom with characteristic features of Majorana fermions, which experimentally are detectable by broad excitation continua due to spin fractionalization. The latter is predicted for the Kitaev spin liquid, an exactly solvable model with bond-dependent interactions on a two-dimensional honeycomb lattice. Here we report on detailed THz experiments in α-RuCl₃, identifying these characteristic fingerprints of Majorana fermions. The continuum intensity decreases and finally vanishes on increasing temperature. It partly overlaps with phonon modes representing characteristic sliding and compression modes of the van der Waals bonded molecular layers.

Majorana fermions are exotic particles, which are identical with their antiparticles. It is still an open question whether they exist in nature as elementary building blocks of matter. Recently, the realization of Majorana fermions as fundamental quasiparticles came into the focus of solid-state physics. Such quasiparticles can be realized in quantum wires based on semiconductor-superconductor hybrid structures [1], via exchange-coupled ferromagnetic atoms [2], or in vortex cores of iron-based superconductors [3]. Besides their importance in correlated quantum matter, considerable interest stems from possible applications in quantum information [4]. Majorana quasiparticles also can be observed in spin liquids: α-RuCl₃, with effective spin ½ moments on a honeycomb lattice, is a prime candidate to host a Kitaev spin-liquid (KSL) ground state. The two-dimensional (2D) Kitaev model [5] with bond-dependent interactions is exactly solvable via the fractionalization of quantum spins into two types of Majorana fermions: $Z_2$ fluxes and itinerant fermions. Details of the Kitaev physics and appropriate models and materials are discussed in recent review articles [6,7].

In scattering experiments, these fractionalized excitations can be observed as characteristic continua up to temperatures comparable to the typical Kitaev exchange of order ~ 100 K. There are detailed theoretical proposals to observe spin fractionalization in the dynamic response via inelastic neutron scattering [8,9,10,11,12], Raman scattering [13,14], optical THz spectroscopy [12,15,16,17], as well as resonant inelastic x-ray-scattering experiments [18]. Indeed, broad continua in α-RuCl₃ were reported in neutron-scattering measurements [19,20,21,22], as well as in Raman [23,24] and THz experiments [25,26,27,28]. As will be discussed later, experimentally the continua appear at low energies and extend up to 30 meV. However, it has to be mentioned that the interpretation of these broad and featureless intensities in terms of fractional excitations of the spin-liquid remains controversial. It was proposed that the observed continua may represent incoherent excitations originating from strong magnetic anharmonicity [6,29] and a model using realistic exchange interactions reproduces the evolution of the dynamical response at finite temperatures and in external magnetic fields [30]. Very recently, the observation of a half-integer thermal quantum Hall effect was interpreted as a hallmark of fractionalization of quantum spins [31] and as the proof that a topological KSL state in α-RuCl₃ indeed is realized.

In this work, we present detailed time-domain THz experiments in a broad frequency and temperature regime on high-quality single crystals of α-RuCl₃. Our results provide experimental evidence for the existence of a continuum with a characteristic temperature dependence, which we interpret in terms of fractionalized spin excitations. In addition, the continuum significantly decreases when passing the structural phase transition from the low-temperature rhomobohedral to the high-temperature monoclinic phase. It signals decreasing importance of Kitaev exchange in the honeycomb lattice with monoclinic distortions. This magnetic continuum partly overlaps with phonon-like excitations, which can be described as characteristic rigid plane shear and compression modes of molecular layers that are only weakly bonded by van der Waals (vdW) forces. These rigid-plane modes are characteristic features of 2D crystal structures and were observed in graphene multilayers and in di-chalcogenides [32,33,34,35].

Systematic THz experiments on α-RuCl₃ have been reported earlier, mainly focusing on the evolution of the low-temperature magnetic scattering [26]. To avoid the influence of phonon excitations and of the structural phase transition, all spectra were normalized to 60 K and the evolution of a continuum was documented below this temperature only. Here we focus on the full spectral range up to room temperature, taking into account that spin fractionalization will dominate the dynamic conductivity at least up to temperatures comparable to the Kitaev exchange of order ~ 100 K.

α-RuCl₃ is a layered compound, with each honeycomb lattice of ruthenium embedded within two hexagonal layers of chlorine. $Ru^{3+}$ ($4d^5$) is octahedrally coordinated by Cl-ions, at room temperature with slight monoclinic distortion, and exhibits an effective spin ½ ground state. The ruthenium ions strongly bonded to adjacent chlorine layers, represent rigid molecular units, which are only weakly connected by vdW forces to each other. Recent experiments on high-quality single crystals converge towards a monoclinic (C2/m) room-temperature structure [36], which transforms via a strongly hysteretic first-order phase transition into a low-temperature rhombohedral phase [27,28, 37]. In the rhombohedral phase with $R\bar{3}$ symmetry, the molecular stacks exhibit an ABC type layering [37]. The phase transition into the monoclinic C2/m structure induces only little change in the geometry of the $RuCl_3$ slabs. There the molecular layers are displaced along the crystallographic a direction yielding slight deviations from the low-temperature three-layer stacking periodicity. The main structural change is an increase in the interlayer spacing [28,37]. The monoclinic C2/m high-temperature phase exhibits a single-layer periodicity only. The effective spin ½ of the ruthenium ions on the honeycomb lattice undergo long-range antiferromagnetic order close to 7 K [28] and exhibit a characteristic zig-zag spin pattern [36].

High-quality α-$RuCl_3$ single crystals were grown by vacuum sublimation. The sample characterization is described in detail in Refs. [21,37] and in the supplementary of Ref. [28]. In the samples that have been investigated in the course of this work, the structural phase transition appeared close to 170 K (on heating) and antiferromagnetic order was established below 6.5 K. The samples for the optical THz experiments had a typical ab surface of 5 × 3 $mm^2$ and a thickness of ~ 1 mm. Time-domain THz transmission experiments were performed with the wave vector of the incident light perpendicular to the crystallographic ab plane using a TPS Spectra 3000 spectrometer. We measured time-domain signals for reference (empty aperture) and samples, from which power spectra were evaluated via Fourier transformation. The transmission spectra show characteristic modulation due to multiple scattering events within the sample. Whenever necessary, we tried to correct for these interference effects by calculating the interference pattern for a given thickness and refractive index of the sample.

Figure 1 documents the temperature dependence of the real part of the dielectric constant ε' and of the real part of the dynamic conductivity σ', as determined by THz spectroscopy for energies between 1 and 14 meV and temperatures between 4.5 and 300 K. Both quantities document that the dipolar strengths of all excitations in this energy regime are very low. This fact becomes clear by the low conductivity [see Fig. 1(b)] and, concomitantly, by the minor contributions of these excitations to the frequency dependence of the dielectric constant [see Fig. 1(a)]. The continuous decrease of the dielectric constant with decreasing energy results from contributions of high-energy phonons (not shown).

At the lowest temperatures, the conductivity spectra [Fig. 1(b)] are dominated by a narrow excitation close to 2.5 meV, a broad continuum around 8.5 meV, and a strong increase beyond 12 meV. The latter results from a weak phonon mode centered close to 15 meV [27, 38]. An additional weak excitation close to 6 meV strongly overlaps with the continuum. The features at the lowest and highest frequencies are barely temperature dependent, while the continuum mode steadily decreases with increasing temperature and finally vanishes beyond 200 K.

First hints that the dynamic conductivity documented in Fig. 1(b) may indicate remnants of Kitaev physics, stem from theoretical work of Bolens et al. [16]. These authors calculated the low-frequency optical conductivity of Kitaev materials assuming an interplay of Hund's coupling, spin-orbit coupling, and crystal-field effects. They predict a gap, followed by a narrow low-frequency peak of magnetic dipolar character, and a broad dominant peak of electric dipolar nature. These theoretical predictions are in qualitative agreement with the low-temperature conductivity shown in Fig. 1(b). However, it is obvious that the weak temperature dependence of the 2.5 meV excitation opposes this interpretation. Focusing on the temperature dependence and assuming that all effects of spin fractionalization should vanish at temperatures far beyond the Kitaev exchange (> 200 K), it seems natural to identify the low-energy (~ 2.5 meV) as well as the high-energy response (> 12 meV) as being due to phonon excitations. The further weak excitation located close to 6 meV, mentioned above, also likely is of phononic nature. In clear contrast, the broad continuum centered around 8.5 meV, which vanishes at high temperatures, indeed could signal fractionalized excitations of the KSL.

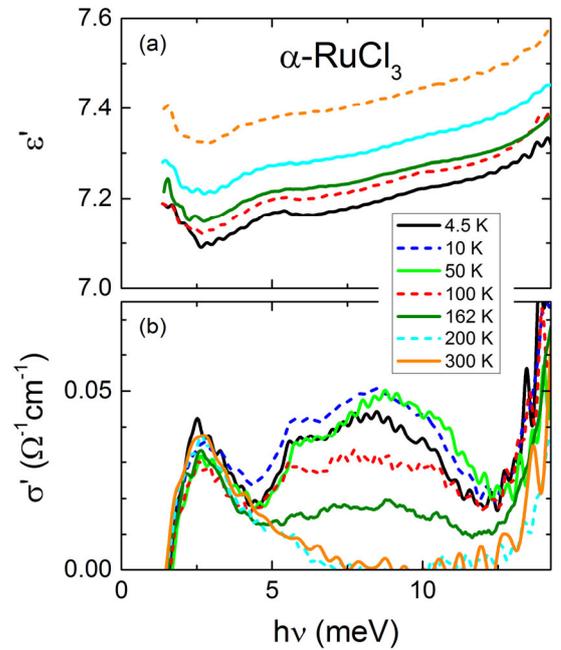

FIG. 1. Energy dependence of (a) the dielectric constant ε' and (b) the dynamic conductivity σ' in α-$RuCl_3$ in the THz regime for a series of temperatures between 4.5 and 300 K. These data have been taken in transmission in normal incidence on heating. All the spectra shown were corrected for multiple scattering effects.

A simple and straightforward experimental proof of the microscopic origin of the excitations under consideration can be provided by a plot of the normalized spectra, which will differentiate between phononic (weak temperature dependence) and magnetic scattering (strong temperature dependence when crossing the magnetic transition). Figure 2 shows the measured transmission at normal incidence at a series of temperatures between 4.5 and 20 K, with all spectra normalized to 7 K, a temperature close to the onset of magnetic order at $T_N$ = 6.5 K. In this plot, phononic excitations will be simply ruled out by their weak variation in this narrow temperature regime and the normalized transmission is dominated by excitations of magnetic origin with significant temperature dependence.

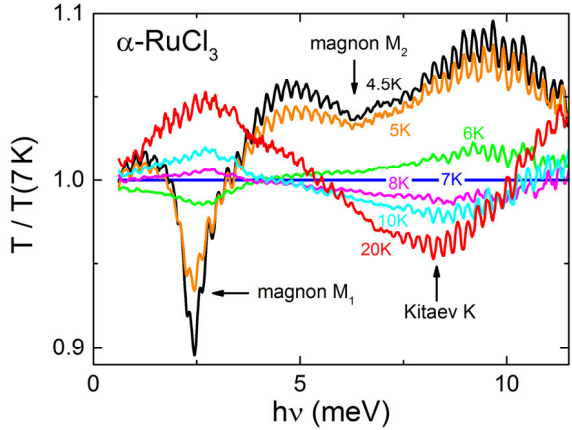

FIG 2. Energy dependence of the transmission in α-RuCl$_3$ in the THz regime, for energies between 1 and 12 meV and temperatures between 4.5 and 20 K. All transmission spectra were normalized to 7 K, a temperature close to the Néel temperature, $T_N \sim 6.5$ K. The arrows M$_1$ and M$_2$ indicate magnon excitations in the antiferromagnetic phase. The arrow K signals the evolution of a continuum due to fractionalized excitations of the Kitaev-type spin liquid in the paramagnetic phase.

The normalized transmission spectra of Fig. 2 impressively document how the magnon modes M$_1$ and M$_2$ evolve below the onset of magnetic order at 6.5 K. Due to the strong phonon response arising at nearly the same frequencies, these magnetic excitations can hardly be identified in the conductivity spectra of Fig. 1(b). They appear close to 2.5 and 6.5 meV and were clearly identified by inelastic neutron scattering [20,39], but also by previous THz experiments [25,26,40,41]. The low-frequency excitation M$_1$ is a one-magnon-response, while the weak and broader high-frequency excitation M$_2$ probably results from two-magnon scattering. For temperatures above the onset of magnetic order, both excitations become fully suppressed. However, a broad scattering continuum K centered around 8.5 meV evolves in the paramagnetic phase, which reduces the transmission in this energy range considerably. This is the first proof that the broad continuum, documented in Fig. 1(b) is not connected to conventional excitations, resulting from long-range magnetic order, but results from fractionalized spin excitations of the KSL. The temperature evolution of the transmission certainly excludes an interpretation in terms of spin fluctuations close to the onset of magnetic order. In the paramagnetic phase, e.g. at 8 K, just above T$_N$, the magnetic modes have completely disappeared, and a continuum evolves at significantly higher energies, documented at 8, 10, and 20 K, and increases on increasing temperatures at least up to 20 K, which corresponds to three-times the Néel ordering temperature of 6.5 K.

To further discriminate between excitations of phononic and magnetic origin, we trace the temperature dependence of the dynamic conductivity σ'. The results as determined close to the peak values of σ'(ω) are documented in Fig. 3. The temperature dependence of the conductivity at these two energies behaves significantly different [please note the different scales in Figs. 3(a) and (b)]. On decreasing temperatures, σ'(T) at 2.4 meV is only weakly temperature dependent and remains close to 0.04 Ω$^{-1}$cm$^{-1}$. There is a small step-like decrease at the structural phase transition and a rather significant enhancement below the onset of magnetic order. This increase signals the appearance of the magnetic mode at 2.5 meV. Under cooling, the conductivity at 8.5 meV continuously increases from zero conductivity (within experimental uncertainty), exhibits a step-like increase at the structural phase transition, and then further increases down to the onset of magnetic order. In the magnetically ordered state, a decrease of the intensity signals a decrease of the dipolar strength of the continuum: Continuum intensity from fractionalized spin excitations is transferred to magnon modes with well-defined spin components.

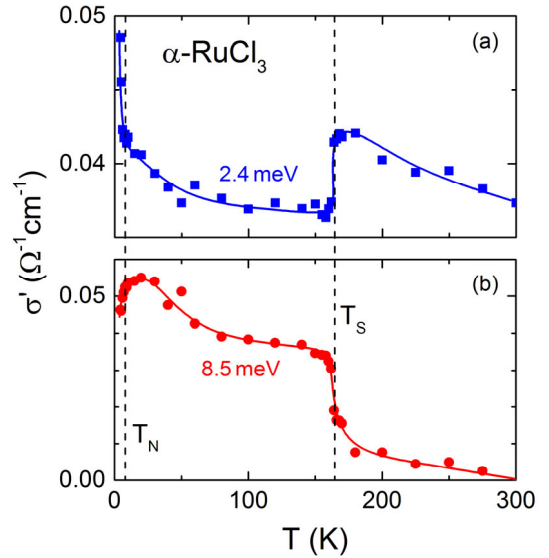

FIG 3. Temperature dependence of dynamic conductivity σ' in α-RuCl$_3$ for energies at (a) 2.4 and (b) 8.5 meV, close to the maxima of the low- and high-energy excitations, respectively. The magnetic (T$_N$) and structural phase-transition temperatures (T$_S$) are indicated by dashed vertical lines. The solid lines are drawn to guide the eye.

In first respect, in the paramagnetic phase the low-frequency mode behaves phonon-like, while the high-frequency continuum indeed pertains to fractionalized excitations, which vanish at a temperature scale larger than the Kitaev interaction (> 200 K). It is worth to mention two further details: The high-frequency conductivity is significantly reduced in the high-temperature monoclinic phase [Fig. 3(b)]. This signals that Kitaev exchange is more dominant in the low-temperature rhombohedral phase. Indeed, the honeycomb lattice of the ruthenium ions is close to ideal in the low-temperature rhombohedral phase and slightly distorted in the high-temperature monoclinic phase [37]. In addition, a minor fraction of the conductivity is transferred from high to low frequencies at the onset of magnetic order. This is compatible with neutron scattering results [20], which show that in the antiferromagnetic phase the continuum is still finite, but is significantly reduced with intensity transfer to magnon modes indicating localized magnetic moments.

Returning to Fig. 1, we conclude that the low-frequency peak at 2.5 meV and the weak 6.5 meV excitation represent phonon-like excitations, while the high-frequency continuum results from fractionalized excitations. What remains to be clarified, is the nature of these phonons. In conventional phonon analysis of tri-halides, these frequencies are much too low for modes within the molecular RuCl$_3$ layers. However, similar to a large variety of two-dimensional lattices, where the molecular stacks are bonded by weak vdW interactions, for the tri-halides one expects low-frequency modes that characterize shear and layer-breathing motions of rigid molecular layers [34]. For a more quantitative analysis, we have fitted the energy-dependent conductivity of α-RuCl$_3$.

Figure 4(a) shows fits of σ'(ω) between 1 and 13 meV for a series of characteristic temperatures. The fits were performed using three excitations in this energy regime and one phonon mode fixed at 15 meV, assuming Lorentzian line shapes (multiplied by frequency to account for the conductivity). Figure 4(b) shows the deconvolution of the fit at 10 K into two narrow and one rather broad excitations. The narrow modes, which exist up to room temperature, represent phonon modes, while the broad line, which vanishes for temperatures > 200 K, characterizes the response of the KSL due to spin fractionalization. In our interpretation, the broad continuum located around 8.5 meV signifies the presence of Majorana fermions. The two phonon modes at 2.5 and 6 meV correspond to rigid-plane shear and breathing modes of molecular layers, respectively. Support for the latter notion comes from recent ab-initio phonon calculations, which predict these excitations close to 3 and 7 meV [42].

In summary, we have detected an absorption continuum in α-RuCl$_3$, which extends from the lowest measured energies of ~ 2 meV up to ~ 15 meV. We identify it with fractionalized spin excitations of the KSL ground state, signifying the presence of Majorana-like quasiparticles. As function of increasing temperature, it continuously decreases and vanishes within experimental uncertainty above 200 K. From the detailed temperature dependence of the continuum intensity (Fig. 3b), we conclude that the step-like decrease of its optical weight, when entering the high-temperature monoclinic phase, signals a significant lowering of the Kitaev exchange in this phase. Indeed, the rhombohedral phase exhibits an almost ideal honeycomb lattice [37]. That, indeed, Kitaev interactions in the monoclinic phase are reduced, compared to competing Heisenberg and off-diagonal exchange, also follows from the fact that the magnetic ordering temperatures of monoclinic crystals are enhanced by a factor of two [36].

Our results compare well with those from neutron-scattering: Banerjee et al. [20] identify a continuum with a maximum close to 5 meV extending up to 15 meV. In the magnetically ordered state, the continuum is reduced and its intensity is transferred to magnon excitations, similar to our findings. Do et al. [21] identify a continuum again extending up to 15 meV, which vanishes roughly at 150 K. In neutron scattering there are no detailed reports of scattering intensities when passing the structural phase transition. In Raman experiments the continuum extends to significantly higher frequencies (~ 30 meV) and survives almost up to room temperature [14,23,24]. Fractionalization of spin degrees of freedom in α-RuCl$_3$ was also derived from heat-capacity experiments [21,42]. Widmann *et al.*, utilizing a phonon correction based on ab-initio calculations, identified a heat capacity anomaly close to 70 K extending up to 150 K [38]. These temperatures are in good agreement with the energies derived from the scattering continuum reported in this THz study.

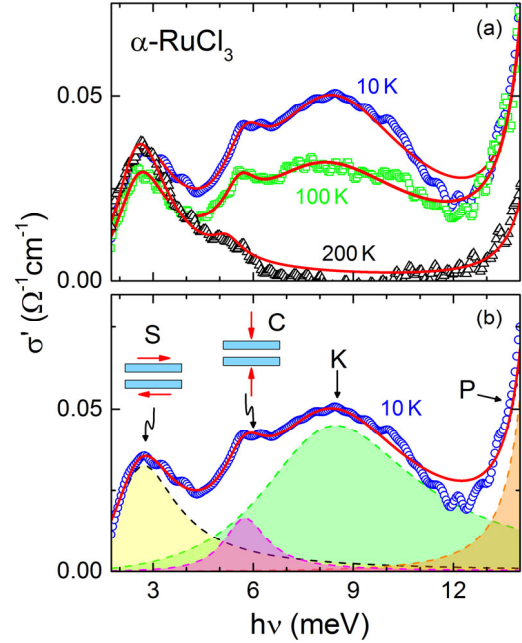

FIG. 4: Analysis of the energy dependent dynamic conductivity in α-RuCl$_3$. (a) Conductivity at a series of temperatures between 10 and 200 K. The solid lines show fits as described in the text. (b) Deconvolution of the conductivity spectrum at 10 K, using three phonon excitations S, C, and P located at 2.5, 6, and 15 meV and a broad continuum K close to 8.5 meV. The phonon modes close to 2.5 and 6 meV represent sliding (S) and compression (C) modes of rigid molecular RuCl$_3$ layers, as schematically indicated in (b).

Finally, we reported on the observation of sliding and breathing modes of the molecular layers, a characteristic feature of 2D layered compounds. Similar modes have been identified in graphene and in di-chalcogenides. In α-RuCl$_3$ these modes are located close to 2.5 and 6 meV. Strictly speaking, these are acoustic modes and according to ab-inito phonon calculations [42] these values correspond to eigenfrequencies at the zone boundary along the crystallographic *c* direction. In THz spectroscopy, these modes probably can be observed via disorder of the stacking or via backfolding of zone-boundary intensities due to the ABC stacking sequences of the rhombohedral phase. Concluding, in this letter we provide a detailed analysis of the temperature dependent THz spectra in α-RuCl$_3$: We find convincing evidence for rigid-layer modes of the molecular units and for a continuum due to Majorana fermions, characteristic excitations of the Kitaev-type spin liquid state.

We thank H. M. Rønnow for bringing the physics of sliding modes in two-dimensional systems to our attention. This work has been partly supported by the Deutsche Forschungsgemeinschaft (DFG) via the Transregional Collaborative Research Center TRR 80.

*Corresponding author: alois.loidl@physik.uni-augsburg.de